\documentclass[12pt]{article}

\usepackage{cite}
\usepackage{subfigure}
\usepackage{multirow}
\usepackage{helvet}
\usepackage{amsmath}
\usepackage{amssymb}
\usepackage{setspace}
\usepackage{setspace}
\usepackage[dvips]{graphicx}
\usepackage{epsfig}

\begin{document}

\titlepage                                                    
\begin{flushright}                                                    
IPPP/13/46  \\
DCPT/13/92 \\                                                                                                       
\end{flushright} 
\vspace*{0.5cm}
\begin{center}                                                    
{\Large \bf A simple form for the low--$x$ generalized parton\\[0.2 cm] distributions in the skewed regime}\\

\vspace*{1cm}
                                                   
L.A. Harland--Lang \\                                                 
                                                   
\vspace*{0.5cm}                                                    
Institute for Particle Physics Phenomenology, University of Durham, Durham, DH1 3LE

\vspace*{1cm}                                                    
                                                    
\begin{abstract}                                                    
\noindent
We show that the generalized parton distributions (GPDFs) in the `skewed' $x \approx \xi \ll 1$ regime can be related in a particularly simple way to the usual diagonal distributions. This follows directly from the Shuvaev transform, but bypasses a direct evaluation of the poorly convergent double integral which this relation contains, thus allowing for a more transparent understanding of the physics involved; it avoids the necessity for any further low--$x$ approximations for the diagonal partons, as well as permitting a clearer application to cases when the GPDFs unintegrated over the parton transverse momentum are required. We consider for illustration the specific examples of the central exclusive production of a Standard Model Higgs boson, and the photoproduction of $J/\psi$ and $\Upsilon$ mesons at the LHC, and show how a careful evaluation of the GPDFs, which our simple results allow, is required to correctly calculate the predicted cross sections.

\end{abstract}                                                        
\vspace*{0.5cm}                                                    
                                                    
\end{center}  

\section{Introduction}

It is well known that suitably defined cross sections in hard hadronic processes can be described in terms of universal diagonal parton distribution functions (PDFs). The inclusive cross section is given by summing over all possible hadronic final states $X$: this is a crucial step in defining the diagonal PDFs, which only depend on the momentum fraction $x$ of the struck parton and the scale $\mu^2$, allowing the process cross section to be written in a factorized form, i.e. as a convolution of the parton level cross section and the corresponding PDFs.

However, for less inclusive observables a treatment in terms of conventional PDFs cannot necessarily be directly applied. One interesting example of this is the class of `elastic' hadronic processes, such as deeply virtual Compton scattering ($\gamma^* p \to \gamma p$), diffractive vector particle production ($\gamma^{(*)} p \to V p$ where $V=\rho, J/\psi, \Upsilon, Z$...), or central exclusive production \linebreak[4]($pp \to p\, +\,X\,+\,p$, where $X=$ Higgs particle, dijets, $\chi_c$...). In this case the processes are described in terms of so--called generalized parton distribution functions (GPDFs), see~\cite{Diehl:2003ny,Belitsky:2005qn} for reviews and a complete list of references. The GPDFs are represented for the gluon case in Fig.~\ref{fig:var}: they depend on the momentum fractions $x_{a,b}=x\pm \xi$ carried by the emitted and absorbed partons, as well as the momentum transfer variable $t=(p-p')^2$ and scale $\mu^2$, and the factorization is written in terms of the parton level amplitude and 
corresponding GPDFs.

Unfortunately, the cross sections for such exclusive processes are generally small, and there are insufficient data to determine these distributions with an accuracy comparable to that of the global parton analyses for the diagonal distributions, although fits do exist, see for example~\cite{Kumericki:2009uq,Sabatie:2012pe,Kroll:2012sm} and references therein. Fortunately, under some reasonable physical assumptions it is possible to determine the GPDFs from the diagonal distributions for the case of small $\xi \ll 1$. The basic idea is that in this regime, the non--zero skewedness should come mainly from the evolution, with the input distributions having $x \gg \xi$, and therefore to good approximation given by the diagonal distributions (for which $x_a=x_b$). This is expressed mathematically in the Shuvaev transform~\cite{Shuvaev:1999fm,Shuvaev:1999ce}, which takes advantage of the observation that the conformal moments of the GPDFs coincide at LO with the usual Mellin moments of the diagonal partons, up to 
corrections of $O(\xi^2)$.  It is a specific case of the more general double distribution~\cite{Mueller:1998fv,Musatov:1999xp,Radyushkin:2000uy} representation of the GPDFs, when the $x^m$--moments of the double distribution $\tilde{f}(x,\alpha ;,\mu)$ have the asymptotic profile function. It has been shown (see~\cite{Musatov:1999xp}) that both of these approaches lead to the same result. We discuss the details of the Shuvaev transform, and its possible limitations, further in Section~\ref{sec:shu}.

A particularly interesting case for which the Shuvaev transform can be applied is that of diffractive processes with large rapidity gaps, for example the central exclusive production (CEP) and photoproduction processes mentioned above. The cross sections for these are written in terms of GPDFs, and typically have $\xi \approx x \ll 1$: for example, at the LHC we have  $\xi \sim 10^{-3}$($10^{-4}$) for diffractive $\Upsilon$($J/\psi$) production and $\xi \sim 10^{-2}$ for central exclusive Higgs production. Recalling that the Shuvaev transform receives corrections of $O(\xi^2)$, these are clearly well within the required kinematic regime. In fact, here the situation is in principle complicated by the fact that the required distributions are the GPDFs {\it unintegrated} over the parton transverse momentum $k_\perp$. However, in the relevant kinematic regime for these processes, it has been shown in~\cite{Kimber:1999xc,Martin:2001ms} that such unintegrated GPDFs can be related back to the integrated ones.

Often in the literature, the full Shuvaev transform, which is given by a quite poorly convergent double integral relation, is not directly used (see for instance~\cite{Khoze:2000cy,Ivanov:2004ax,Kowalski:2006hc,Martin:2007sb,Maciula:2011iv,Jones:2013pga} for some representative examples). Rather, assuming that at low--$x$ the diagonal PDFs behave as a simple power $\sim x^{-\lambda}$, an analytic expression for $x=\xi$, as in~\cite{Shuvaev:1999ce}, is for simplicity taken. While in some situations this gives a sufficiently accurate result, this is not always the case. This can in particular be true when the unintegrated GPDFs are required, where the scale dependence of the diagonal gluons, and therefore the power $\lambda$ enters directly, requiring a more careful treatment. In this paper, we show that the Shuvaev transform for $x=\xi$ can be recast in a different form, with the GPDFs given by a nicely convergent and simple single integral, which can be used for an arbitrary input diagonal PDF. As well as 
having the 
benefit of simplicity, this will allow us to show clearly how a proper treatment of the scale dependence of the GPDFs can be quite important when calculating for example the cross sections for various exclusive processes.

The outline of this paper is as follows. In Section~\ref{sec:shu} we present some quantitative details and further discussion of the Shuvaev transform. In Section~\ref{sec:an} we present the new analytic forms for the GPDFs in the $x=\xi \ll1$ region; the principle results are given by (\ref{yint}) and (\ref{yintq}). We demonstrate that these results can also be used to a good degree of accuracy away from this exact limit, i.e. for $x \approx \xi$.  In Section~\ref{sec:CEP} we apply these to the specific case of Higgs Boson CEP at the LHC, and show how a commonly used approximate formulae for the unintegrated GPDFs can underestimate the predicted cross sections. We also explore more generally the region of validity of the small--$x$ approximate form for the GPDFs. In Section~\ref{sec:quark} we consider the case of vector meson photoproduction at the LHC, finding the effect to be smaller for the specific examples of $J/\psi$ and $\Upsilon$ production, although not necessarily negligible. Finally in Section~\ref{sec:conc} 
we conclude.

\section{The Shuvaev Transform}\label{sec:shu}

\begin{figure}
\begin{center}
\includegraphics[scale=1.2]{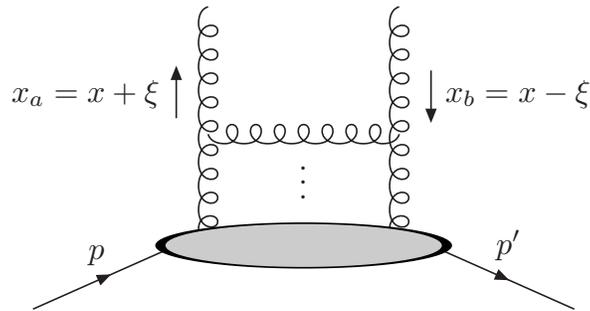}
\caption{A schematic diagram showing the variables for the off-diagonal parton distribution $H (x, \xi)$, for the gluon case.}
\label{fig:var}
\end{center}
\end{figure}

The generalized parton distributions (GPDFs) are denoted~\cite{Ji:1996ek,Ji:1996nm,Ji:1998pc} by $H_{q,g}(x,\xi,\mu^2,t)$, for partons emitted and absorbed at a scale $\mu$, where $\xi$ is the skewing parameter, as shown in Fig.~\ref{fig:var}, with $-1<x<1$. For definiteness we will take $\xi >0$. In general the GPDFs also depend on the squared momentum transfer $t=(p-p')^2$ between the incoming and outgoing hadrons, however, as we have $|t|\ll \mu^2$, this dependence is typically assumed to factorize from the longitudinal momentum dependence as a form factor $F(t)$, and we will assume this to be the case in what follows, omitting the $t$ dependence for simplicity. In the limit that $\xi \to 0$ (that is, $x_a=x_b=x$ in Fig.~\ref{fig:var}), the GPDFs must reduce, from the optical theorem, to the usual diagonal distributions
\begin{align} \label{eq:aa}
H_q (x, 0) & = \left\{
\begin{aligned}
& q (x) \quad\;\; &{\rm for} \quad x > 0\;, \\
-& \bar{q} (- x) \quad &{\rm for} \quad x < 0\;,
\end{aligned}\right.
\\[0.2cm] \label{hgdiag}
H_g (x, 0) &=  x  g (x)\;,
\end{align}
where for simplicity we will leave the scale dependence implicit in what follows. We refer the reader to~\cite{Ji:1998pc,Radyushkin:1997ki,GolecBiernat:1998ja,Diehl:2003ny,Belitsky:2005qn} for detailed treatments.

More generally, for $\xi \ll 1$, in the space--like region $|x|>\xi$, it can be shown that the GPDFs can be determined from the diagonal ones via the Shuvaev transform~\cite{Shuvaev:1999fm,Shuvaev:1999ce}. In particular, the evolution of the GPDFs may be viewed as the renormalization of their conformal moments~\cite{Ohrndorf:1981qv}. 
\begin{equation}\label{conf}
O_N^i(\xi)=\int {\rm d}x\,\mathcal{R}_n^i(x_a,x_b)H_i(x,\xi)\;, 
\end{equation}
where $i=q,g$, the $x_{a,b}$ are defined in Fig.~\ref{fig:var}, and the $\mathcal{R}_n^i$ are known functions, see e.g.~\cite{Shuvaev:1999fm}. For $\xi\ll 1$, these moments can be shown to reduce at LO to the usual Mellin moments of the diagonal distributions, up to corrections of $O(\xi^2)$. Thus, provided the GPDFs can be determined from their conformal moments (\ref{conf}), then the distributions can be related in turn to the diagonal ones. This is achieved by the Shuvaev transform, and it can in particular be shown that~\cite{Shuvaev:1999ce}
\begin{align}\label{eq:shuvq}
H_q (x, \xi) &=  \int_{-1}^1 \: {\rm d}x^\prime \left [ \frac{2}{\pi} \: {\rm Im} \: \int_0^1 
\: \frac{{\rm d}s}{y (s) \: \sqrt{1 - y(s) x^\prime}} \right ] \:
\frac{{\rm d}}{{\rm d}x^\prime} \left ( 
\frac{q (x^\prime)}{| x^\prime |} \right )\;, \\[0.2cm]
\label{eq:shuvg}
H_g (x, \xi) &=  \int_{-1}^1 \: {\rm d}x^\prime \left [ \frac{2}{\pi} 
\: {\rm Im} \: \int_0^1 \: \frac{{\rm d}s (x + \xi (1 - 2s))}{y (s) \: \sqrt{1 - y (s) x^\prime}} 
\right ] \: \frac{{\rm d}}{{\rm d}x^\prime} \left ( \frac{g (x^\prime)}{| x^\prime |} \right ) \:,
\end{align}
where any corrections to these formulae are of order $O(\xi^2)$, and
\begin{equation}
\label{eq:a200}
y(s)=\frac{4s(1-s)}{x+\xi(1-2s)}\,.
\end{equation}
Assuming that the diagonal PDFs have the low--$x$ behaviour
\begin{equation}
x q (x) \; = \; N_q \: x^{- \lambda_q}, \quad\quad xg (x) \; = \; N_g \: x^{- \lambda_g},
\label{eq:pdfpower}
\end{equation}
then (\ref{eq:shuvq}, \ref{eq:shuvg}) are to good approximation given by
\begin{equation}
\label{eq:shuvappr}
\tilde{H}_i (x, \xi)  =  N_i \: \frac{\Gamma \left ( \lambda + \frac{5}{2} \right 
)}{\Gamma (\lambda + 2)} \: \frac{2}{\sqrt{\pi}} \: \int_0^1 \: {\rm d}s \left [ x 
+ \xi (1 - 2s) \right ]^p \: \left [ \frac{4s (1 - s)}{x + \xi (1 - 2s)} \right ]^{\lambda_i + 
1} \; ,
\end{equation}
where $i=q,g$ and $p=0 (1)$ for quarks (gluons). We will often consider in this paper the GPDF in the `skewed' regime, i.e. evaluated at $x_a=x$, $x_b=0$. In this case, it is useful to take the ratio of this to the diagonal distribution at momentum fraction $x$. With this in mind we can define
\begin{equation}
R_i  \equiv  \frac{H_i (x/2, x/2)}{H_i(x, 0)}\;.
\label{rgeq}
\end{equation}
We can use (\ref{eq:shuvappr}) to write
\begin{equation}
\tilde{R}_i \; \equiv \; \frac{\tilde{H} (x/2, x/2)}{H(x, 0)}  =  \frac{2^{2\lambda+3}}{\sqrt{\pi}}
\frac{\Gamma(\lambda + 5/2)}{\Gamma(\lambda + 3 + p)} \;.
\label{eq:Rtildeanalytic}
\end{equation} 
This approximate form was first written down in~\cite{Shuvaev:1999ce}.

Finally, we note that in the literature the validity of the Shuvaev transform has been called into doubt~\cite{Diehl:2007zu}, as in particular any singularities in the right hand complex $N$--plane of the conformal moments $O_N$ can cause additional $O(\xi/x)$ corrections and therefore make the transform inapplicable in practice (in particular when $x \sim \xi $). However, as discussed in~\cite{Martin:2009zzb}, such singularities cannot be generated by the evolution of the GPDFs, and must therefore be present in the input distributions. Under the natural assumption that these low--$x$ input distributions exhibit Regge--like behaviour, they will also contain no such singularities. The Shuvaev prescription is also supported by the good NLO and NNLO fit to HERA DVCS data in~\cite{Kumericki:2009uq}. Thus, while it cannot be proved from first principles, it is both physically motivated and of practical use. We do not consider this question any further in this paper, but rather refer the reader to~\cite{Martin:
2009zzb} and references therein for more discussion.

\section{Analytic form}\label{sec:an}

The double integral (\ref{eq:shuvq},\ref{eq:shuvg}) of the Shuvaev transform is poorly convergent, making the numerical integration quite computationally intensive. On the other hand, assuming (\ref{eq:pdfpower}) to be a good approximation, we can use (\ref{eq:Rtildeanalytic}) for the case $x = \xi$, without performing any integration. A comparison of this approximation with the `exact' result is performed in~\cite{Martin:2009zzb}, and is also considered in Section~\ref{sec:CEP} below. Alternatively, in~\cite{rginter} grid files for different $x,\xi$ values and PDF choices are given, see~\cite{Martin:2009zzb} for more details.

\begin{figure}
\begin{center}
\includegraphics[scale=0.7]{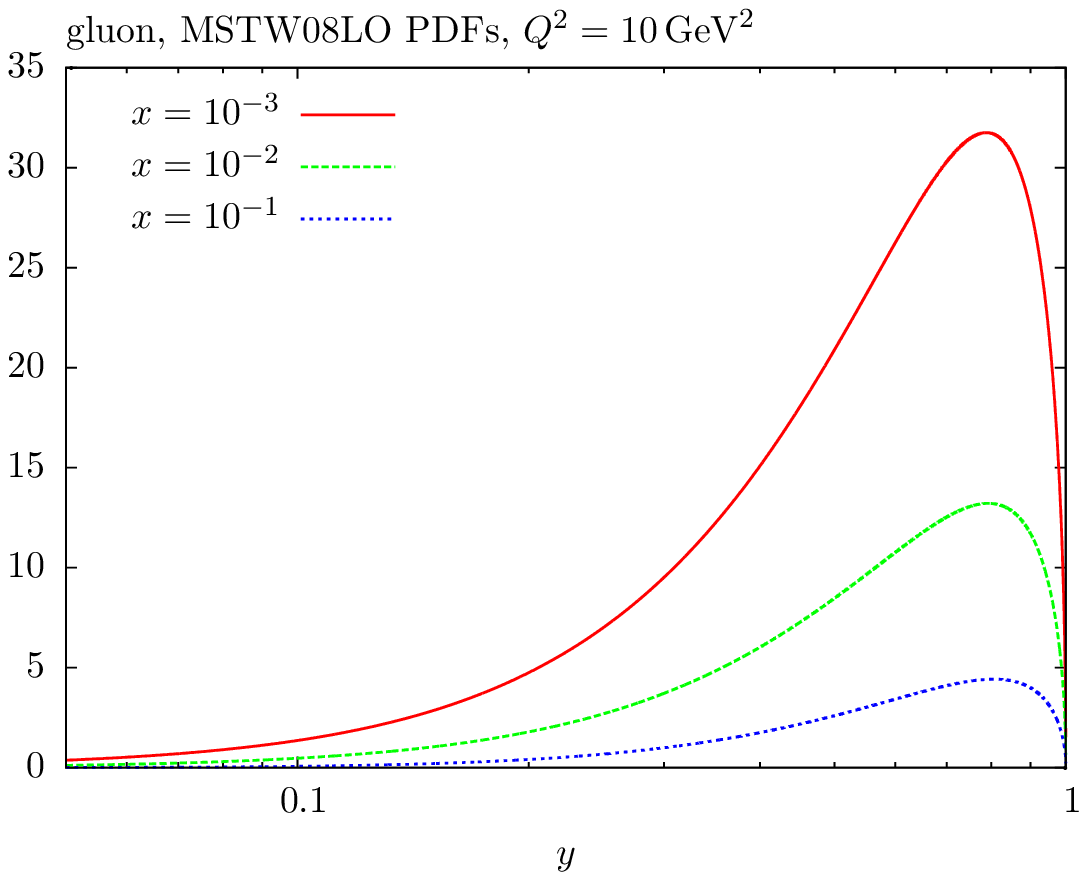}\qquad
\includegraphics[scale=0.7]{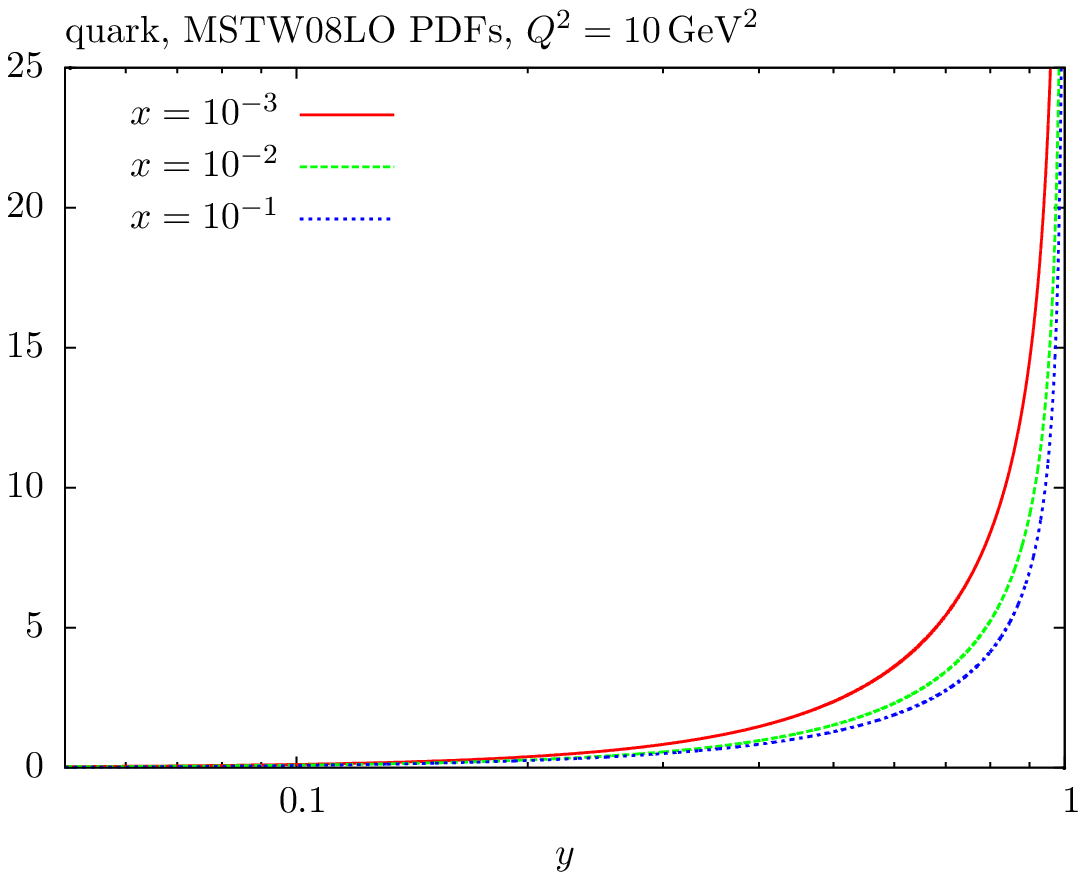}
\caption{Integrands of (\ref{yint}, \ref{yintq}), for different $x$ values. MSTW08L0 PDFs~\cite{Martin:2009iq} are used, at the scale $Q^2=10\,{\rm GeV}^2$. In the quark case the sum of $u$, $d$ and $s$ PDFs are taken, and the integrand is multiplied by an additional factor of $x$.}\label{int}
\end{center}
\end{figure}

Below, we will show that for $x=\xi$, the Shuvaev transform (\ref{eq:shuvq}, \ref{eq:shuvg}) can be written in a very simple form. Considering the gluon case, if we write (\ref{eq:shuvg}) for $x=\xi$ as
\begin{equation}\label{hg}
H_g(x,x)\; = \; \frac{2x^2}{\pi}\;\int_{x/2}^1 \: {\rm d}x^\prime\; I_s(x,x^\prime) \;\frac{{\rm d}}{{\rm d}x^\prime} \left ( \frac{g (x^\prime)}{x^\prime }\right)\;.
\end{equation}
then we find that the integral $I_s(x,x')$ can be performed analytically
\begin{align}\label{ints}
I_s(x,x^\prime) &\equiv {\rm Im} \: \int_0^1 \: {\rm d}s \;\frac{(1-s)}{s\: (1 - 2sx^\prime/x)^{1/2}} \;,\\ \label{ints1}
&=-2\;{\rm arctan}\left[(b-1)^{1/2}\right]+2\,\frac{(b-1)^{1/2}}{b}\;,
\end{align}
where $b=2x^\prime/x$. Using this, and integrating (\ref{hg}) by parts, we find that the surface term vanishes, and we get
 \begin{align}\label{yint}
H_g\left(\frac{x}{2},\frac{x}{2},Q^2\right)\; &= \frac{4x}{\pi} \int_{x/4}^1 \;{\rm d}y \; y^{1/2}(1-y)^{1/2}\,g\left(\frac{x}{4y},Q^2\right)\;,  \\[0.3cm]   \label{yintq}
 H_q\left(\frac{x}{2},\frac{x}{2},Q^2\right)\; &= \left\{
 \begin{aligned}
 &\frac{2}{\pi} \int_{x/4}^1 \;{\rm d}y \; y^{1/2}(1-y)^{-1/2}\,q\left(\frac{x}{4y},Q^2\right) \quad x>0\;,\\
&\frac{2}{\pi} \int_{x/4}^1 \;{\rm d}y \; y^{1/2}(1-y)^{-1/2}\,\overline{q}\left(\frac{x}{4y},Q^2\right) \quad x<0\;,
\end{aligned} \right.
 \end{align}
where $H_g$ is symmetric in $x$. We have relabelled $x \to x/2$ for the sake of comparison with (\ref{eq:Rtildeanalytic}), and reintroduced the explicit scale dependence for clarity. We also show the result for the quark GPDF, which follows from a similar derivation to the gluon case. It can readily be shown that using the small $x$ assumption of (\ref{eq:pdfpower}) in the above expressions reproduces the result (\ref{eq:Rtildeanalytic}), as it must\footnote{In fact, (\ref{yint}, \ref{yintq}) and (\ref{eq:Rtildeanalytic}) are only equivalent when we take the lower limit $x/2 \to 0$ in (\ref{yint}, \ref{yintq}). However, as we are in the $x\ll 1$ regime, and observing the form of the integrand in Fig.~\ref{int}, which are strongly peaked towards $y=1$, it is clear that this is a very good approximation. This point was as also discussed in~\cite{Martin:2009zzb}.}. 

In Fig.~\ref{int} we show the integrands of (\ref{yint},\ref{yintq}), for a range of $x$ values, taking MSTW08LO PDFs~\cite{Martin:2009iq} at scale $Q^2=10\,{\rm GeV}^2$ for illustration. In the quark case we multiply for illustration by an additional factor of $x$: recalling (\ref{rgeq}) and (\ref{eq:aa}), this will give a clearer picture of the size of $R_q$ at different $x$ values. We can see that in both cases the integrand is dominated by the $y\sim 1$ region, that is with the argument of the diagonal PDFs peaked at $x/4$. Thus, at low $x$, the GPDFs in the $x=\xi$ regime are largely independent of the form of the diagonal PDFs at high $x$, lending support to the application of the low--$x$ approximation (\ref{eq:pdfpower},\ref{eq:Rtildeanalytic}). While in the gluon case, the integral is nicely convergent, in the quark case, the integrand is strongly peaked\footnote{With a simple change of variables $u \sim (1-y)^{1/2}$, this integral can be recast into a numerically more manageable form.} at $y=1$, 
leading to in general a much larger ratio $R_q$ (\ref{rgeq}), see for example~\cite{Martin:2009zzb} for some discussion of this. For the sake of brevity, we will not consider the quark GPDF any further in what follows.

\begin{figure}
\begin{center}
\includegraphics[scale=0.7]{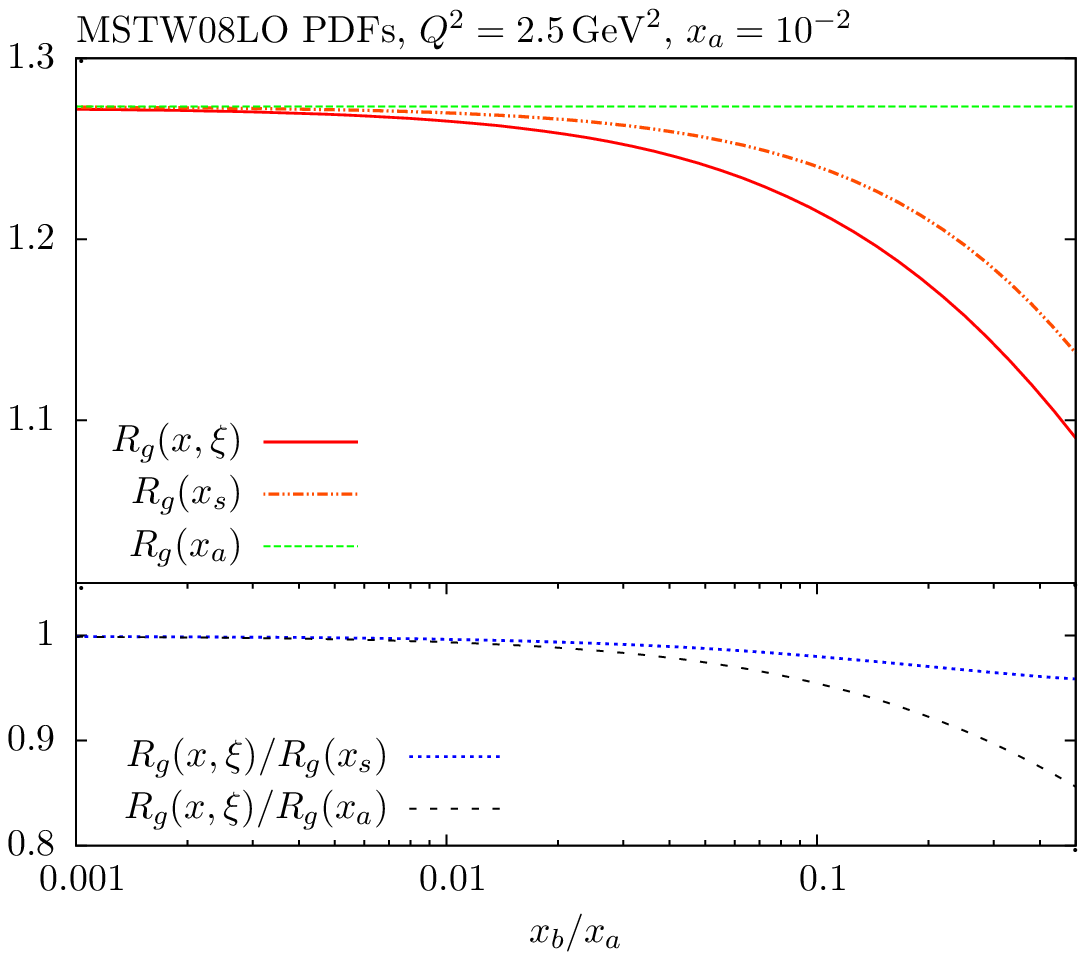}
\includegraphics[scale=0.7]{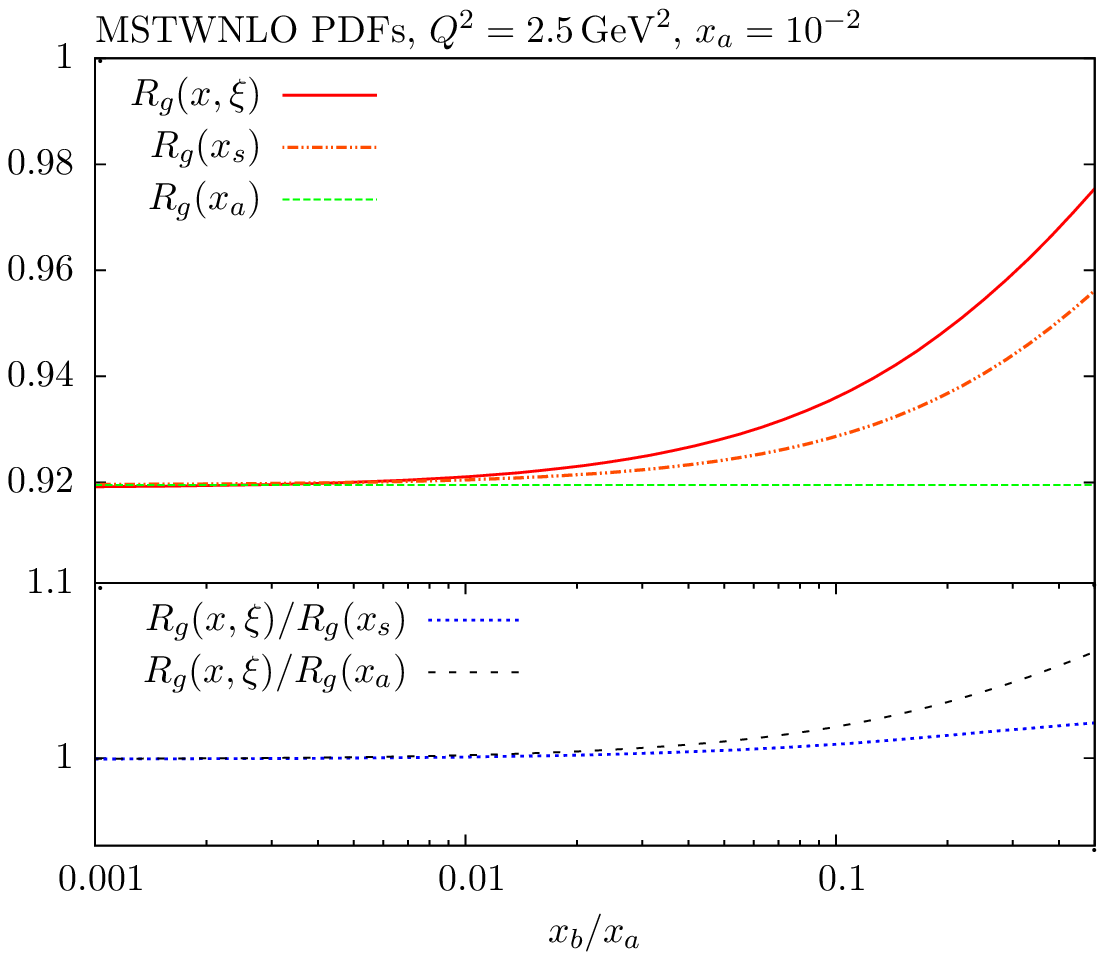}
\caption{Values for the ratio $R_g(x)=H_g(x/2,x/2)/H(x_a,0)$, calculated using (\ref{yint}), for $x=x_a$ and $x=x_s \equiv (x_a+x_b)$, with $x_a=10^{-2}$ and for different values of $x_b$, where the kinematic variables are as defined in Fig.~\ref{fig:var}. Also shown is the full result, $R_g(x,\xi)$, calculated using (\ref{eq:shuvg}).}
\label{fig:rgsxi}
\end{center}
\end{figure}

Although (\ref{yint}, \ref{yintq}) are only strictly valid for $x=\xi$, more generally we may expect, provided $x\approx \xi$, to be able to use these expressions to a good degree of approximation. To explore this possibility, we show in Fig.~\ref{fig:rgsxi} the $R_g$ factor defined in (\ref{rgeq}), calculated using (\ref{yint}), as well as the complete result of (\ref{eq:shuvg}), for the case that $x \neq \xi$, that is $x_b \neq 0$, with the variables defined as in Fig.~\ref{fig:var}. We take fixed $x_a = 10^{-2}, 10^{-3}$ and scale $Q^2= 2.5 \,{\rm GeV}^2$, and use LO and NLO MSTW08 PDFs~\cite{Martin:2009iq}. We can see clearly that these converge very closely for $x_b \ll x_a$, as we would expect, but as $x_b$ increases the results deviate somewhat: while the $R_g$ calculated using (\ref{yint}) with $x=x_a$ is independent of $x_b$, the value found using (\ref{eq:shuvg}) approaches unity as $x_b \to x_a$, as it must. We find a similar result for different PDF sets, while there is some tendency for the 
deviation to increase with $Q^2$, see also the discussion in~\cite{Martin:2009zzb}. For $x_b \gtrsim 0.1\, x_a$, there is already a $\sim 10\%$ deviation for the LO PDFs, and so some care may be needed. For the NLO PDFs the deviation is smaller, although at larger $Q^2$, where the $R_g$ factor becomes positive, this is not necessarily true. Interestingly, if instead of taking the argument $x=x_a$ in (\ref{yint}) we instead consider the argument $x=(x_a+x_b)$, i.e. corresponding to the variable $x$ in the GPDF $H(x,\xi$), we find a much closer matching between this and the full result, using (\ref{eq:shuvg}), even up to quite large values of $x_b\sim x_a$. However, commonly the precise value of the variable $x_b$ may not be known: this will be true in the following sections, when we consider the CEP and photoproduction processes. In this case, we can see that provided $x_b \lesssim 0.1\, x_a$, then taking $x\approx \xi$ and using (\ref{yint}) is valid at the level of a few percent, for the sort of scales, 
$Q^2$, that are relevant to these processes.

Thus we have shown that the GPDFs in the $x \approx \xi$ regime are related in a very simple way to the diagonal PDFs, integrated over the range $[x/4,1]$, with the $\sim x/4$ region giving the dominant contribution. In the following sections we will consider some phenomenological applications of these formulae, and show how their simple form can help clarify some issues associated with these.

\section{The CEP of a Standard Model Higgs boson}\label{sec:CEP}

\begin{figure}
\begin{center}
\includegraphics[scale=0.8,angle=90,trim= 0 0 0 0]{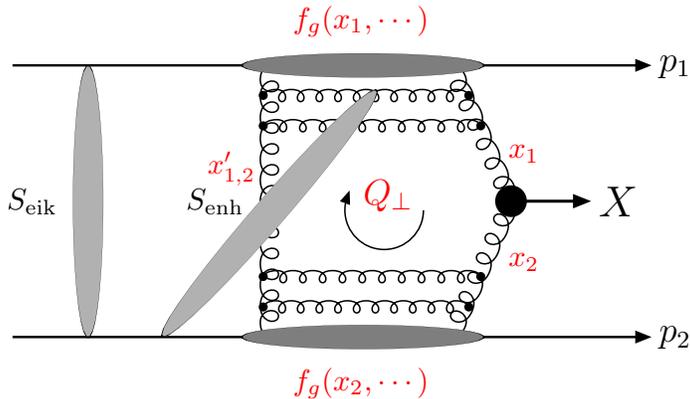}
\caption{The perturbative mechanism for the exclusive process $pp \to p\,+\, X \, +\, p$.}
\label{fig:pCp}
\end{center}
\end{figure} 
 
We will consider central exclusive production (CEP) processes of the type
\begin{equation}\label{exc}
pp({\bar p}) \to p+X+p({\bar p})\;,
\end{equation}
within the `Durham' pQCD--based model (see~\cite{Khoze:2001xm,Albrow:2010yb,HarlandLang:2013jf} for reviews and references), as represented in Fig.~\ref{fig:pCp}, where the coupling of this two--gluon $t$--channel state to the proton (anti--proton) is related to the gluon GPDF. As the $x$ values probed are generally quite low (for example for Higgs production at the LHC we have $x \sim 10^{-2}$), the Shuvaev transform can be used to good approximation, up to small corrections of $O(x^2)$. More specifically, it can be readily shown that the momentum fraction of the screening gluon $x_{1,2}' \sim Q_\perp/\sqrt{s}$, which does not couple to the hard process\footnote{This is unrelated to the integration variable $x'$ in (\ref{hg}). Using the variables of Fig.~\ref{fig:var}, we have $x_{1,2}=x_a$, $x_{1,2}'=x_b$.} is much lower than the momentum fractions $x_{1,2} \sim M_X/\sqrt{s}$ of the active gluons, in the physically relevant $Q_\perp \ll M_X$ regime: for the Higgs case $\langle Q_\perp\rangle$ is $O({\rm GeV})$ and $M_X=M_h\approx 126$ GeV, and so $x' \sim 0.01 \,x$. In the notation of Fig.~\ref{fig:var}, we have $x \approx \xi \ll 1$ and, recalling Fig.~\ref{fig:rgsxi}, we may therefore use (\ref{yint}) to calculate the relevant GPDF to good accuracy.

More specifically, we recall that the perturbative CEP amplitude can be written as~\cite{Khoze:1997dr,Khoze:2004yb,HarlandLang:2012qz}
\begin{equation}\label{bt}
T=\pi^2 \int \frac{{\rm d}^2 Q_\perp}{Q_\perp^6}\,f_g(x_1,x_1', Q_\perp^2,\mu^2)f_g(x_2,x_2',Q_\perp^2,\mu^2)\, \mathcal{M}(gg\to X) \; ,
\end{equation}
where for simplicity we will consider throughout this section the limit that the outgoing proton $p_{\perp}=0$ : this approximation will not affect the conclusions which follow. Here, $\mathcal{M}(gg \to X)$ is the colour--averaged, normalised sub--amplitude for the $gg \to X$ process
\begin{equation}\label{Vnorm}
\mathcal{M}(gg\to X)\equiv -\frac{2}{M_X^2}\frac{1}{N_C^2-1}\sum_{a,b}\delta^{ab}Q_{\perp}^\mu Q_{\perp}^\nu V_{\mu\nu}^{ab} \; ,
\end{equation}
where $M_X$ is the central object mass, $a$, $b$ are the gluon colour indices, and $V_{\mu\nu}^{ab}$ is the $gg \to X$ vertex. We take $\mu=M_X/2$ for the factorization scale. Taking the example of Standard Model Higgs boson production, the CEP amplitude (\ref{bt}) is given by
\begin{equation}\label{bth}
 T^{\rm Higgs}=A\pi^3 \int \frac{{\rm d} Q_\perp^2}{Q_\perp^4}\,f_g(x_1,x_1', Q_\perp^2,\mu^2)f_g(x_2,x_2',Q_\perp^2,\mu^2) \; ,
\end{equation}
where $A$ is a constant given in~\cite{Khoze:2000cy}. The $f_g$'s in (\ref{bt}) are the skewed gluon densities of the proton, unintegrated over the gluon transverse momentum, and corresponding to the $x' \ll x$ limit. They are related to the  (integrated) GPDF via~\cite{Coughlin:2009tr,Martin:1997wy}
\begin{align}\nonumber
f_g(x,x',Q_\perp^2,\mu^2)&=\; \frac{\partial}{\partial \ln(Q_\perp^2)} \left[ H_g\left(\frac{x}{2},\frac{x}{2};Q_\perp^2\right) \sqrt{T(Q_\perp,\mu^2)} \right]\;,\\ \label{fgskew}
&=\; \frac{\partial}{\partial \ln(Q_\perp^2)} \left[ R_g\left(xg(x,Q_\perp^2)\right) \sqrt{T(Q_\perp,\mu^2)} \right]\;.
\end{align}
where $R_g$ is defined in (\ref{rgeq}), which we have introduced to make contact with previous section, and $T$ is the Sudakov factor which ensures that the active gluon does not emit additional real partons in the course of the evolution up to the hard scale $\mu$, so that the rapidity gaps survive. It is given by
\begin{equation}\label{ts}
T(Q_\perp^2,\mu^2)={\rm exp} \bigg(-\int_{Q_\perp^2}^{\mu^2} \frac{{\rm d} {\bf k}_\perp^2}{{\bf k}_\perp^2}\frac{\alpha_s(k_\perp^2)}{2\pi} \int_{0}^{1-\Delta} \bigg[ z P_{gg}(z) + \sum_{q} P_{qg}(z) \bigg]{\rm d}z \bigg) \; .
\end{equation}
with~\cite{Coughlin:2009tr}\footnote{This updated prescription for the $z$ cutoff is used in all papers from~\cite{HarlandLang:2010ep} onwards by the authors.}
\begin{equation}
 \Delta=\frac{k_\perp}{M_X}\;.
\end{equation}
Using (\ref{yint}) we may then readily evaluate (\ref{fgskew}) to calculate the CEP amplitude (\ref{bt}). However, commonly in the literature, two approximations are made (see for instance~\cite{Khoze:2000cy,Ivanov:2004ax,Kowalski:2006hc,Martin:2007sb,Maciula:2011iv,Jones:2013pga} for some representative examples of this). Firstly, any scale dependence of the factor $R_g$ is ignored: that is, the scale dependence of the diagonal and generalized gluon PDFs (\ref{yint}) are assumed to be the same. Secondly, the value of $R_g$ is often found by assuming that the gluon density exhibits the low--$x$ behaviour of (\ref{eq:pdfpower}), and fitting the power $\lambda_g$. In this case we may write (\ref{fgskew}) as
\begin{equation}\label{fgskewap}
f_g(x,x',Q_\perp^2,\mu^2)\approx \tilde{R}_g\frac{\partial}{\partial \ln(Q_\perp^2)} \left[ xg(x,Q_\perp^2) \sqrt{T(Q_\perp,\mu^2)} \right]\;,
\end{equation}
where $\tilde{R}_g$ is given by (\ref{eq:Rtildeanalytic}). While these assumptions have the benefit of simplifying the calculation, avoiding the computationally expensive integration of (\ref{eq:shuvg}), their reliability is certainly not guaranteed. Indeed, from (\ref{bt}) we can see that the CEP cross section will depend on the GPDF to the fourth power, and so some care is needed. Using the simple form (\ref{yint}), we can evaluate (\ref{fgskew}) and test the validity of these approximations.

\begin{figure}
\begin{center}
\includegraphics[scale=0.7]{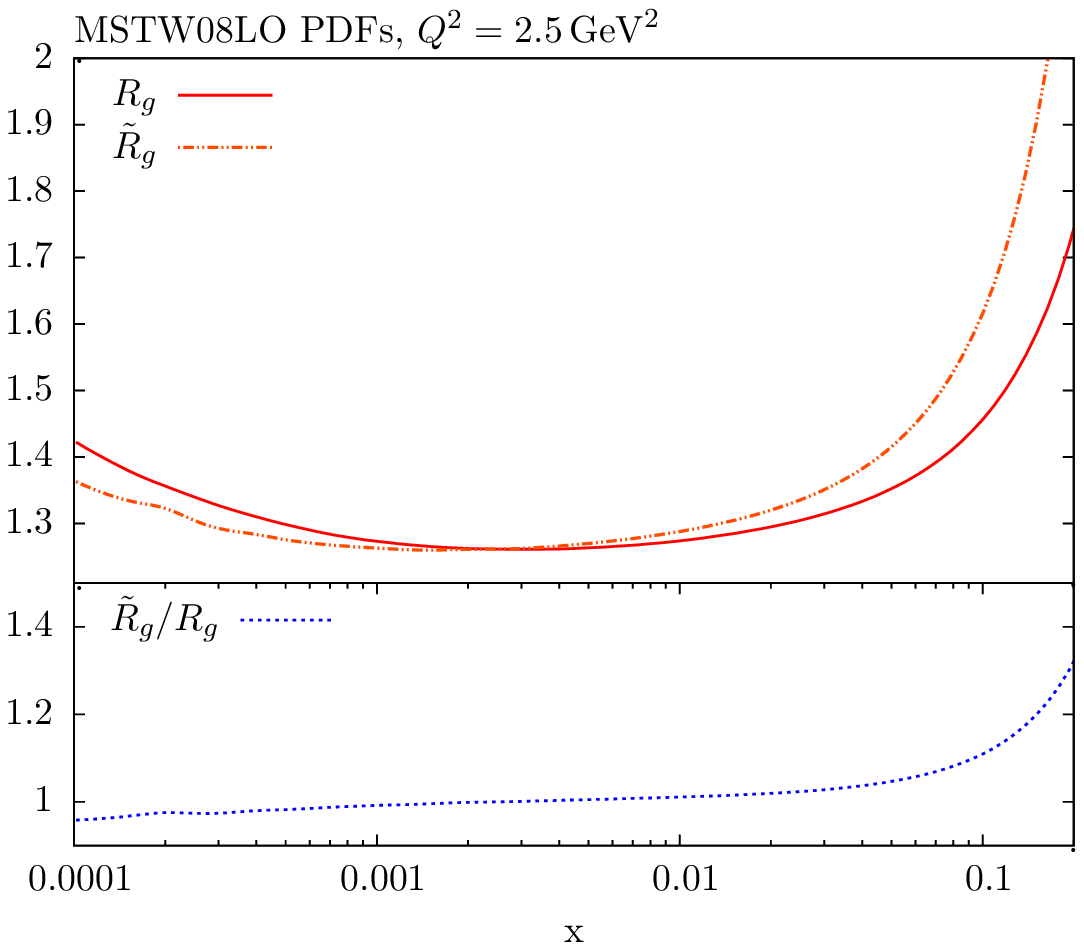}\qquad
\includegraphics[scale=0.7]{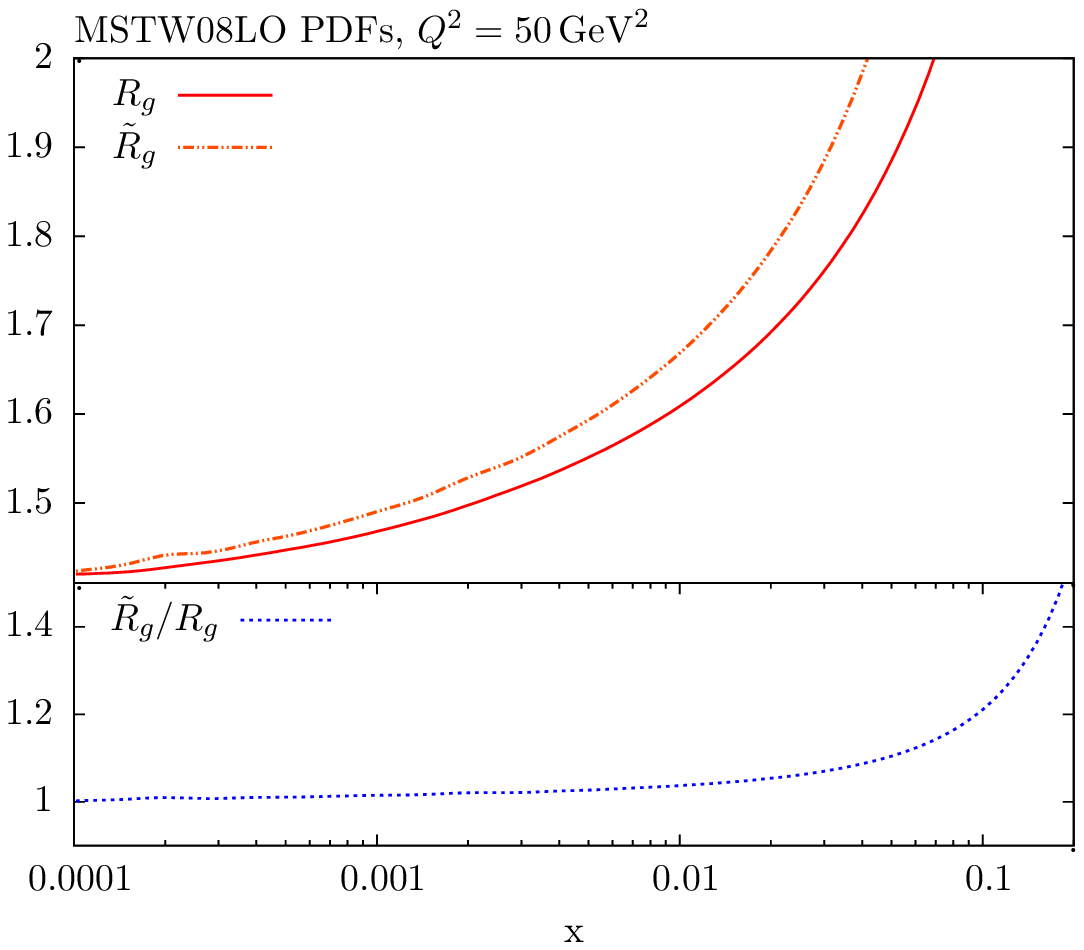}
\caption{`Exact' and `approximate' expressions for the ratio $H_g(x/2,x/2)/H(x,0)$, $R_g$ and $\tilde{R}_g$, calculated using (\ref{rgeq}) with $H_g(x/2,x/2)$ given by (\ref{yint}), and using (\ref{eq:Rtildeanalytic}), respectively. MSTW08LO PDFs~\cite{Martin:2009iq} are used, with scales $Q^2=2.5,\,50\, {\rm GeV}^2$.}
\label{fig:rgs1}
\end{center}
\end{figure}

The question of the latter approximation, in which the gluon density is assumed to exhibit the low--$x$ behaviours (\ref{eq:pdfpower}) has already been considered in the literature, see~\cite{Martin:2009zzb} and references therein for more details. In Fig.~\ref{fig:rgs1} we show the `exact' and `approximate' expressions, $R_g$ and $\tilde{R}_g$. While the `exact' value $R_g$ is calculated\footnote{We note that using the grid files described in~\cite{Martin:2009zzb}, which are calculted using the integral (\ref{eq:shuvg}), give a very similar result to this, and throughout the following sections when (\ref{yint}) is used, as we would expect. However, such files are only available for a limited number of PDF sets, and our approach, using the easily integrated expression in (\ref{yint}), bypasses the need for these in the $x\approx \xi\ll 1$ regime.} using (\ref{rgeq}) with $H_g(x/2,x/2)$ given by (\ref{yint}), the `approximate' value $\tilde{R}_g$ is calculated using (\ref{eq:Rtildeanalytic}), that is, assuming the low--$x$ form of (\ref{eq:pdfpower}). We take the MSTW08LO PDFs~\cite{Martin:2009iq} for two representative choices of scale $Q^2=2.5, 50\,{\rm GeV}^2$: the following conclusions remain essentially unchanged for other choices of PDF. We can see that in general there is quite a good agreement between the two expressions, at 
the percent level, provided the $x$ value is sufficiently small. We note that it is in general not enough that the PDF exhibits the power--
like behaviour of (\ref{eq:pdfpower}) at the $x$ value of interest. We can see from (\ref{yint}) that the generalized gluon density, and therefore $R_g$, is given by an integral over the interval $[x/4,1]$, although as observed in Section~\ref{sec:an} (see Fig.~\ref{int}), the integrand is dominantly peaked towards $x/4$. The approximate expression (\ref{eq:Rtildeanalytic}) assumes this power--like behaviour over the entire range, with a constant power $\lambda_g$. Thus, even in the low--$x$ region there can be some difference between $R_g$ and $\tilde{R}_g$, as we can see in Fig.~\ref{fig:rgs1} (left). As $x$ increases, the value of $R_g$ becomes more sensitive to the high--$x$ region, where such a simple power--like behaviour cannot be justified, with the approximate expression $\tilde{R}_g$ becoming artificially large. Provided we are at sufficiently low $x \lesssim 0.05$ this approximation is nonetheless a very good one, but in the intermediate region of $x\sim 0.1$, where $x$ may still be considered 
`small', this is less clear (although the Shuvaev transform, valid up to corrections of $O(x^2)\sim 1\%$ may still be reliably applied). There may for example be some sensitivity to this region when the particle $X$ is produced at forward rapidity. Our expressions (\ref{yint}), (\ref{yintq}), which make no assumptions about the behaviour of the diagonal PDF, avoid such an issue.

\begin{figure}
\begin{center}
\includegraphics[scale=0.65]{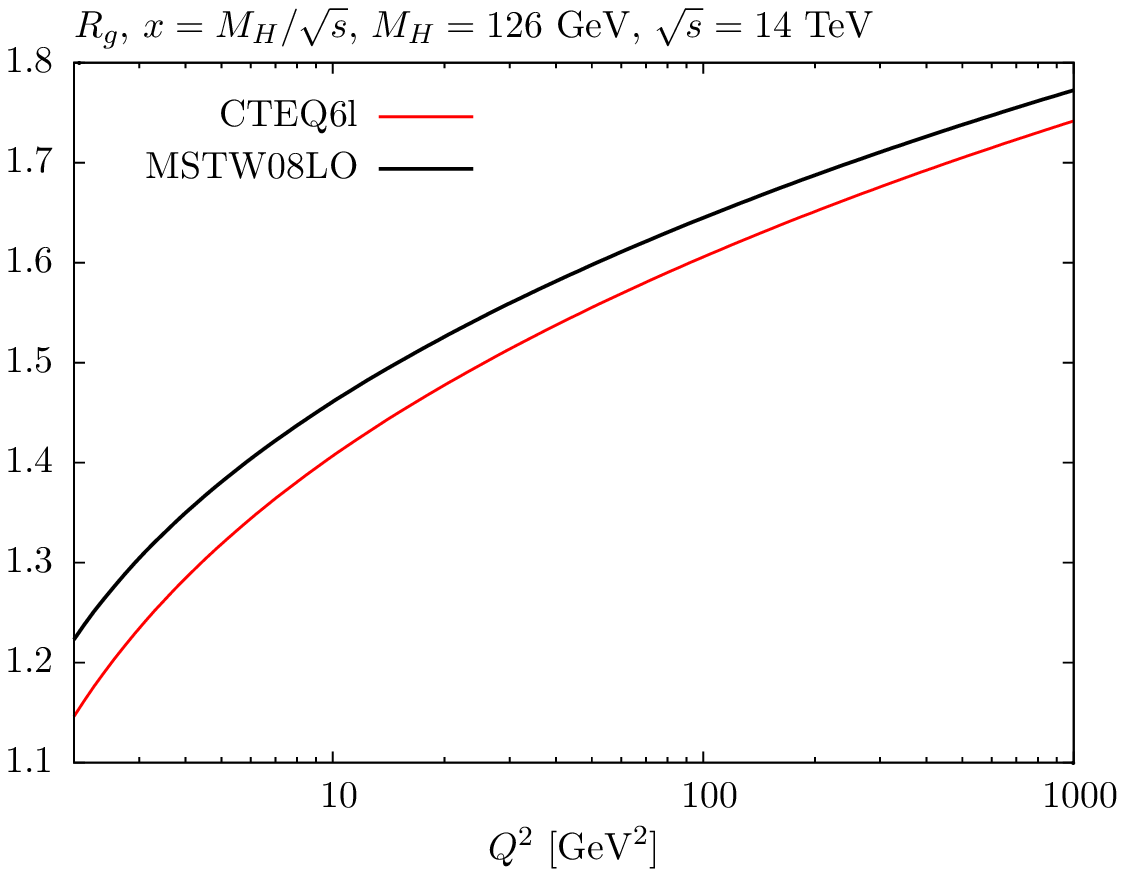}\qquad
\includegraphics[scale=0.65]{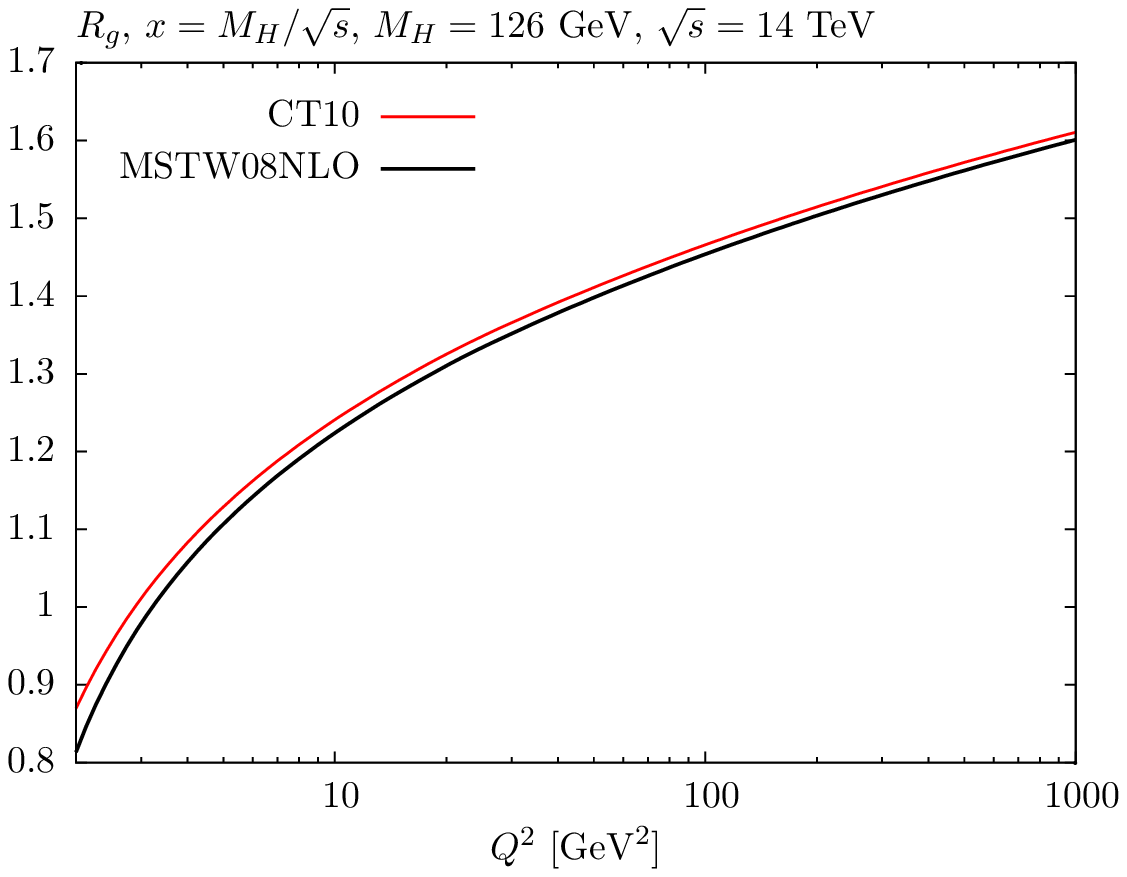}
\caption{The ratio $R_g=H_g(x/2,x/2)/H(x,0)$, with $H_g(x/2,x/2)$ calculated using (\ref{yint}) directly, as a function of the PDF scale $Q^2$. MSTW08LO~\cite{Martin:2009iq} and CTEQ6L~\cite{Pumplin:2002vw} PDFs are used, with $x=M_H/\sqrt{s}$, $M_H=126$ GeV and $\sqrt{s}=14$ TeV.}
\label{fig:rgs}
\end{center}
\end{figure}

\begin{figure}
\begin{center}
\includegraphics[scale=0.65]{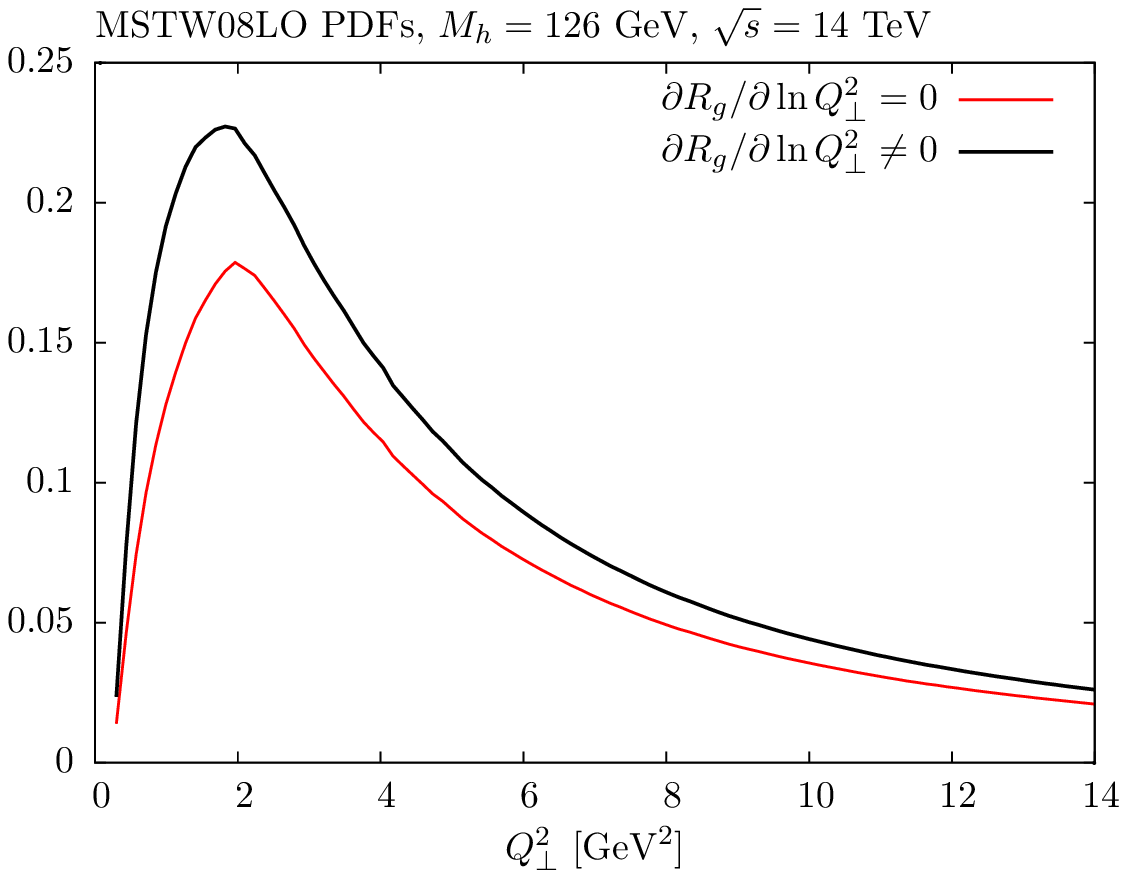}\qquad
\includegraphics[scale=0.65]{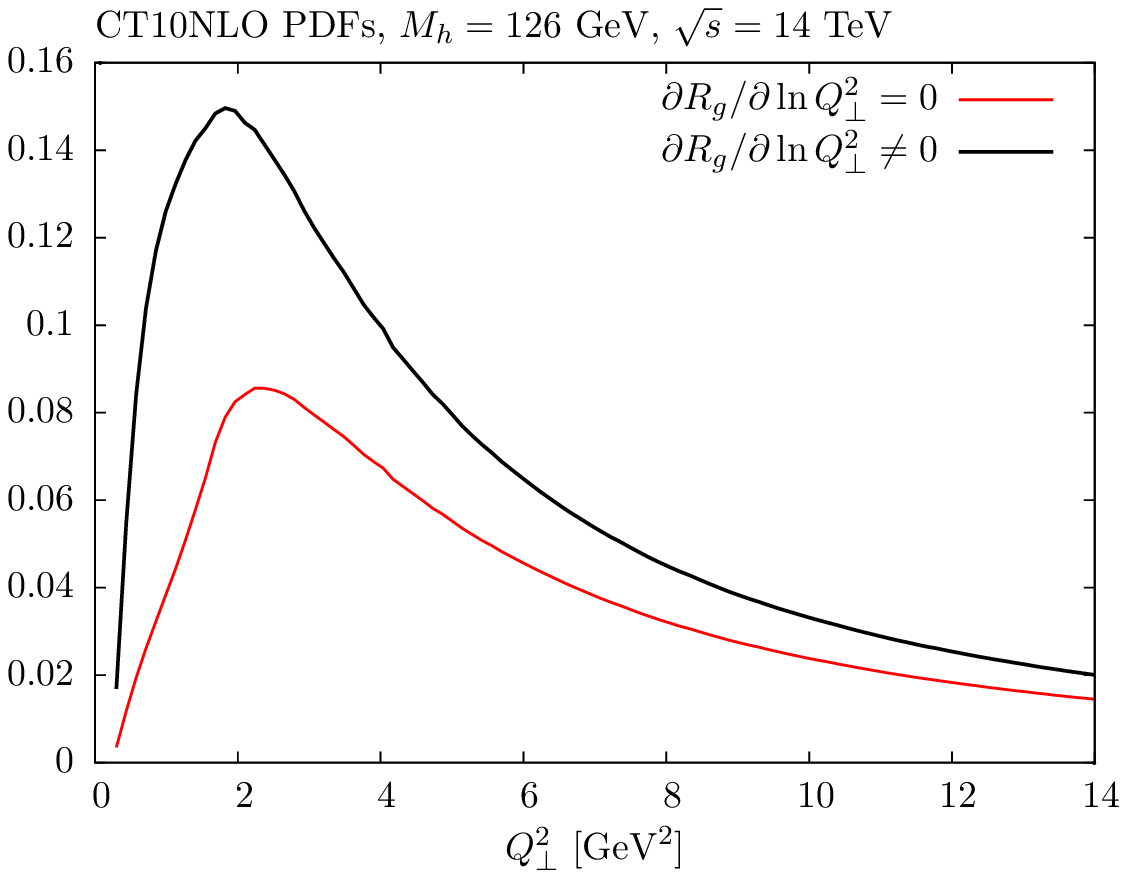}
\caption{The integrand of the CEP amplitude (\ref{bth}) for the production of a $M_H=126$ GeV Higgs Boson, at $\sqrt{s}=14$ TeV, as a function of the gluon loop momentum squared $Q_\perp^2$, with the $R_g$ factor taken inside and outside the differential, as in (\ref{fgskew}) and (\ref{fgskewap}), that is, with the $Q_\perp$ dependence $\partial R_g/\partial \ln Q_\perp^2$ included and excluded, respectively. In both cases the value of $R_g$ is found using (\ref{yint}).}
\label{fig:omt}
\end{center}
\end{figure}

We now turn to the former approximation discussed above, that is, the scale dependence of the factor $R_g$. In Fig.~\ref{fig:rgs} we show $R_g$, calculated using (\ref{yint}), as a function of the scale $Q^2$. We can see that it displays some non--negligible dependence on $Q^2$, bringing this approximation into question (we can also see, as in~\cite{Martin:2009zzb}, that using the NLO diagonal partons, the GPDFs are actually suppressed relative to these at lower scales, as here the effective power $\lambda_g$ in (\ref{eq:pdfpower}) becomes negative). The relative size of this contribution will depend on the value of $\partial R_g/\partial \ln Q_\perp^2$, but also on the other terms which come when the differential (\ref{fgskew}) is expanded out. This therefore requires a precise numerical comparison, using the correct, $Q_\perp$ dependent expression for $R_g$. The simple form of (\ref{yint}) which we have derived in this paper allows this to be done with ease. 

In Fig.~\ref{fig:omt}, we show the integrand of the CEP amplitude (\ref{bth}) for the production of the Standard Model Higgs Boson at $\sqrt{s}=14$ TeV as a function of the gluon loop momentum squared\footnote{In the low $Q_\perp$ region to which the diagonal PDF sets do not necessarily extend, it is necessary to perform some extrapolation, e.g. by freezing the gluon anomalous dimension $\gamma$ below some low scale $Q_0$, so that $xg(x,Q^2)\sim (Q^2)^\gamma$. As the dominant part of the integrand comes from higher scales $Q_\perp>Q_0$, we find that the CEP amplitude is largely independent of the details of such an extrapolation.} $Q_\perp^2$, using MSTW08LO~\cite{Martin:2009iq} and CT10 PDFs~\cite{Lai:2010vv} as a representative choice (other PDF sets give similar results). We show both the integrand using the full result (\ref{fgskew}) for the skewed PDFs, that is with the $R_g$ factor taken inside the differential, and with the approximation of (\ref{fgskewap}), that is with this $Q_\perp$ dependence neglected. The difference is quite big, with the approximation of (\ref{fgskewap}), which omits the positive contribution from $\partial R_g/\partial \ln Q^2$, underestimating the size of the CEP amplitude. In Table~\ref{pdft} we show cross sections for the CEP of the SM Higgs Boson at $\sqrt{s}=14$ TeV, using a range of PDF sets. As before, these are calculated with and without the $R_g$ factor included inside the differential (\ref{fgskew}). The numerical importance of calculating the $R_g$ factor precisely is clear, with the approximation of (\ref{fgskewap}) tending to underestimate the CEP cross section by a factor of up to $2$. In general, for lower object masses $M_X$ and/or higher $\sqrt{s}$, where the $x$ values probed are lower, we find that the numerical effect is less pronounced, but still non--negligible. We also show the cross section for the case that the approximation (\ref{eq:Rtildeanalytic}) is used to calculate $\tilde{R}_g$, but with this included inside the differential (\ref{fgskew}). In this case, although the positive contribution from $\partial \tilde{R}_g/\partial \ln Q^2$ is included, this approximation tends to overestimate the cross section somewhat, as we might expect from Fig.~\ref{fig:rgs1}.

\renewcommand{\arraystretch}{1.2}
\begin{table}
\begin{center}
\begin{tabular}{|l|c|c|c|c|c|}
\hline
&MSTW08LO&CTEQ6L&GJR08LO(FF)&CT10&NNPDF2.1\\
\hline
$\partial R_g/\partial Q^2 =0$&0.83&1.15&1.94&0.27&0.19\\ 
\hline
$\partial R_g/\partial Q^2 \neq0$&1.39&1.91&2.66&0.56&0.40\\
\hline
fit&1.22&1.92&2.52&0.65&0.50\\
\hline
$\partial \widetilde{R}_g/\partial Q^2 \neq0$&1.57&2.59&3.22&0.74&0.51\\
\hline
\end{tabular}
\caption{Cross sections in fb for Higgs Boson ($M_H=126$ GeV) CEP at $\sqrt{s}=14$ TeV, integrated over the rapidity interval $-2.5<y_H<2.5$, using a range of PDF sets~\cite{Martin:2009iq,Pumplin:2002vw,Gluck:2007ck,Lai:2010vv,Ball:2010de}. These are calculated with the $R_g$ factor taken inside and outside the differential, as in (\ref{fgskew}) and (\ref{fgskewap}), that is, with the $Q_\perp$ dependence $\partial R_g/\partial \ln Q_\perp^2$ included and excluded, respectively. In both cases the value of $R_g$ is found using (\ref{yint}). We also show the result of the fit (\ref{fgfit}) of~\cite{Martin:2001ms}, with $\tilde{R}_g$ calculated using (\ref{eq:Rtildeanalytic}). Finally, the cross section with $\tilde{R}_g$ calculated using the approximation (\ref{eq:Rtildeanalytic}), but included inside the differential as in (\ref{fgskew}), that is with the $Q_\perp^2$ dependence $\partial \tilde{R}_g/\partial \ln Q_\perp^2$ included, is shown.}\label{pdft}
\end{center}
\end{table}
\renewcommand{\arraystretch}{1}

For completeness, we also show the result of the phenomenological fit of~\cite{Martin:2001ms}, modified slightly to account for the results of~\cite{Coughlin:2009tr} for the limit $\Delta$ entering the Sudakov factor $z$ integral, see (\ref{ts}). Explicitly this gives
\begin{align}\nonumber
 f_g(x,x',Q_\perp^2,\mu^2)&= \sqrt{T} \bigg[ \tilde{R}_g \frac{\partial x g(x,Q_\perp^2)}{\partial \ln Q_\perp^2}+xg(x,Q_\perp^2)\frac{N_c \alpha_s}{2\pi}\left(\ln \bigg(\frac{M_X}{Q_\perp}\right)+\\ \label{fgfit}
 &+1.2\frac{\mu^2}{\mu^2+Q_\perp^2}\bigg) +5\frac{\alpha_s}{2\pi}\left(xu_{\rm val}(x,Q_\perp^2)+xd_{\rm val}(x,Q_\perp^2)\right)\bigg]\;,
\end{align}
where $\tilde{R}_g$ is calculated using (\ref{eq:Rtildeanalytic}). This or a very similar form is used to produce the Higgs cross section predictions in~\cite{HarlandLang:2013jf} and by the paper authors in~\cite{HarlandLang:2010ep} onwards. The form of (\ref{fgfit}) is fitted to the full result, which is derived by explicitly substituting (\ref{eq:shuvg}) into the $Q_\perp$ evolution equation for the GPDFs. Such a fit will not reproduce the full result completely, and may be less reliable for other PDF choices (the fit was performed using MRST99 PDFs). Nonetheless, it approximately includes the effect of the $\partial R_g/\partial \ln Q_\perp^2$ term in (\ref{fgskew}), and so we can see in Table~\ref{pdft} that the Higgs cross sections calculated using this fit reproduce to quite good approximation the complete result. Thus, the previous predictions of~\cite{HarlandLang:2010ep} do not need to be significantly modified. Nonetheless, it is clear that the form of (\ref{fgskew}) for the skewed PDFs should in 
general be used, and our result (\ref{yint}) allows this to be done in a very simple way.

\section{The photoproduction of heavy quarkonia}\label{sec:quark}

\begin{figure}
\begin{center}
\includegraphics[scale=1.2]{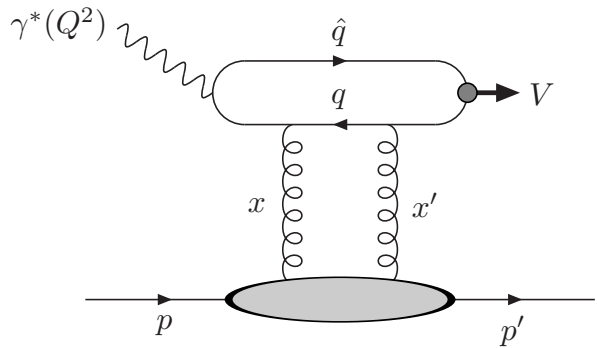}
\caption{Schematic picture of pQCD mechanism for the photoproduction of heavy quarkonia. Quark (anti--quark) momenta flow from left to right.}
\label{fig:jpsi}
\end{center}
\end{figure}

Another process in which the gluon GPDF plays a crucial role is the photoproduction of heavy quarkonia
\begin{equation}
 \gamma^* p(\overline{p}) \to V p(\overline{p})\;.
\end{equation}
Due to the presence of the hard scale set by the quark mass, this can be modelled perturbatively, as shown in Fig.~\ref{fig:jpsi}. The colour singlet $Vp$ interaction is mediated by a two--gluon exchange in the $t$--channel, with the coupling of this to the proton related to the gluon GPDF. To first approximation, the photoproduction cross section is given by~\cite{Ryskin:1992ui}
\begin{equation}\label{jpsilo}
\frac{{\rm d}\sigma}{{\rm d}t}(\gamma^*p \to V p)|_{t=0}=\frac{\Gamma_{ee}M_V^3\pi^3}{48\alpha}\left[\frac{\alpha_s(\overline{Q}^2)}{\overline{Q}^2}xg(x,\overline{Q}^2)\right]^2\left(1+\frac{Q^2}{M_V^2}\right)\;,
\end{equation}
where $\Gamma_{ee}$ is the width of the $V \to e^+e^-$ decay. However such an expression ignores the contributions which come beyond the leading log from an explicit integration over the gluon $k_\perp$, as well as any `skewedness' of the gluon GPDFs\footnote{For simplicity we do not account for other corrections here, due to, e.g., $c\overline{c}$ rescattering, relativistic corrections and the real part of the amplitude, which should also in general be considered, see~\cite{Ryskin:1995hz,Martin:1999rn}}  i.e. it assumes that $R_g=1$. A more careful treatment (see for example~\cite{Martin:2007sb}) shows that to include these corrections we should make the replacement in (\ref{jpsilo})
\begin{align}\nonumber
 \frac{xg(x,\overline{Q}^2)}{\overline{Q}^2} &\to \int_{Q_0^2}^{(W^2-M_V^2)/4} \frac{ {\rm d} k_\perp^2}{(\overline{Q}^2+k_\perp^2)k_\perp^2}\frac{\partial [H_g(x,\xi,k_\perp^2) \sqrt{T(k_\perp^2,\mu^2)}]}{\partial \ln k_\perp^2}\;,\\ \label{jpsikt}
 &=\int_{Q_0^2}^{(W^2-M_V^2)/4} \frac{ {\rm d} k_\perp^2}{(\overline{Q}^2+k_\perp^2)k_\perp^2}\frac{\partial [R_g \left(xg(x,k_\perp^2)\right) \sqrt{T(k_\perp^2,\mu^2)}]}{\partial \ln k_\perp^2}\;,
\end{align}
where
\begin{equation}
 \overline{Q}^2=(Q^2+M_V^2)/4\;,
\end{equation}
where $T(k_\perp^2,\mu^2)$ is the Sudakov factor, defined in (\ref{ts}), and the factorization scale $\mu=M_X/2$. The momentum fractions of the $t$--channel gluons are given by
\begin{equation}\label{xxp}
  x'=\frac{M_{q\overline{q}}^2-M_V^2+k_\perp^2}{W^2+Q^2}\;, \qquad \qquad x=\frac{M_{q\overline{q}}^2+Q^2+ k_\perp^2}{W^2+Q^2}\;,
\end{equation}
where $M_{q\overline{q}}^2=(q+\hat{q})^2$ is the mass of the intermediate $q\overline{q}$ system. This depends on both the gluon $k_\perp$ and the distribution of the quark momenta within the meson, but for $k_\perp^2 \ll M_V^2$, and in the non--relativistic approximation for the meson wave function, we have $M_V \to M_{q\overline{q}}$ and so $x' \to 0$, in which case we are in precisely the $x=\xi$ regime described in Section~\ref{sec:an}, with the gluon GPDF given by (\ref{yint}). More generally, provided the meson mass and/or photon $Q^2$ is sufficiently large, we will have $x \approx \xi$ to sufficiently good approximation that this approach can still be used. 
In this case, comparing this expression with (\ref{fgskew}) we can see that again, a careful treatment of the scale dependence of the GPDF is needed to evaluate the cross section.

\begin{table}
\begin{center}
\begin{tabular}{|l|c|c|c|c|c|}
\hline
&MSTW08LO&CTEQ6L&GJR08NLO(FF)&CT10&Fit\\
\hline
$x=10^{-2}$&1.41&1.35&1.28&1.64&1.29\\
\hline
$x=10^{-3}$&1.13&1.23&1.17&1.43&1.23\\
\hline
$x=10^{-4}$&0.88&1.16&1.13&1.16&1.19\\
\hline
\end{tabular}
\caption{Ratio $\sigma(\partial R_g/\partial Q^2 \neq 0)/\sigma(\partial R_g/\partial Q^2 = 0)$ using different PDF sets for the $\gamma p \to J/\psi p$ photoproduction cross section at different values of the gluon $x$. Both cross sections are calculated using (\ref{jpsikt}), with (\ref{yint}) used to calculate the $R_g$ factor, but in the case of the denominator, the additional approximation as in (\ref{fgskewap}) is made. The `Fit' refers to the NLO parameterisation of~\cite{Martin:2007sb}, which has been extracted from $J/\psi$ photoproduction data.}\label{pdftjpsi}
\end{center}
\end{table}

\begin{table}
\begin{center}
\begin{tabular}{|l|c|c|c|c|c|}
\hline
&MSTW08LO&CTEQ6L&GJR08NLO(FF)&CT10&Fit\\
\hline
$x=10^{-2}$&1.44&1.49&1.37&1.73&1.33\\
\hline
$x=10^{-3}$&1.18&1.27&1.21&1.36&1.25\\
\hline
$x=10^{-4}$&0.99&1.18&1.15&1.15&1.19\\
\hline
\end{tabular}
\caption{Ratio $\sigma(\partial R_g/\partial Q^2 \neq 0)/\sigma(\partial R_g/\partial Q^2 = 0)$ using different PDF sets for the $\gamma p \to \Upsilon p$ photoproduction cross section at different values of the gluon $x$. Both cross sections are calculated using (\ref{jpsikt}), with (\ref{yint}) used to calculate the $R_g$ factor, but in the case of the denominator, the additional approximation as in (\ref{fgskewap}) is made. The `Fit' refers to the NLO parameterisation of~\cite{Martin:2007sb}, which has been extracted from $J/\psi$ photoproduction data.}\label{pdftjpsi1}
\end{center}
\end{table}

In previous estimates (see e.g.~\cite{Jones:2013pga,Martin:2007sb,Martin:1999wb} and references therein), the same approximation (\ref{fgskewap}) as in Section~\ref{sec:CEP} has in general been used. However, as we have seen above, there may be some non--negligible corrections to this, which a precise numerical evaluation of (\ref{jpsikt}) can clarify. In Tables~\ref{pdftjpsi} and~\ref{pdftjpsi1} we therefore show the ratio of the cross section for $\gamma p \to J/\psi(\Upsilon) p$ photoproduction, calculated using (\ref{jpsikt}), to the cross section calculated using the same expression, but with the approximation of (\ref{fgskewap}) made, omitting the scale dependence of the $R_g$ factor. We show results for a range of PDF sets and $x$ values, as well as using the NLO fit of~\cite{Martin:2007sb}. We can see that, in particular for higher $x$ values, the more precise expression (\ref{jpsikt}) predicts in general a somewhat larger cross section and that the $x$ dependence as well as overall normalization is 
affected. These results therefore indicate that in general applying the approximation (\ref{fgskewap}), as has been done in previous studies, may not be completely valid. On the other hand at lower $x$, where the data tends to lie, the changes in Tables~\ref{pdftjpsi} and~\ref{pdftjpsi1} are quite small.

Finally, we note that some care is needed when treating the low $k_\perp$ region of the integral (\ref{jpsikt}). Here, the form of the diagonal PDFs is not known and a perturbative treatment cannot necessarily be trusted. Although the presence of the Sudakov factor ensures the result is finite, if the hard scale $\overline{Q}^2$ is not sufficiently large, the cross section may display some sensitivity to this region. In for example~\cite{Martin:2007sb} a cut--off  $k_\perp > Q_0$ is imposed on the integral in (\ref{jpsikt}), with an additional constant piece accounting for the $k_\perp<Q_0$ region; in the current paper we prefer to omit such a cut--off, and integrate (\ref{jpsikt}) down to $k_\perp \sim \Lambda_{\rm QCD}$, with a smooth extrapolation performed for the diagonal gluons at low scale. If a perturbative treatment is to be applicable, then the final result should not be too dependent on the choice of $Q_0$ or details of the extrapolation, and conversely any sensitivity to this is indicative of an 
intrinsic uncertainty in the perturbative treatment. For the case of $\Upsilon$ production, we find that the predicted cross sections are not too sensitive to this, with the ratios presented in Table~\ref{pdftjpsi1} only changing by at most a few percent when the form of the extrapolation or cutoff $Q_0$ is changed between reasonable choices. However, as the scale decreases, for example in the case of $J/\psi$ photoproduction (with photon $Q^2 \approx 0$), this sensitivity increases, with the results of Table~\ref{pdftjpsi} changing by up to $\sim 10 \%$, although the overall trends with $x$ remain.
Moreover, it may be also be the case, see (\ref{xxp}), that we are not at sufficiently low $x' \ll x$ for (\ref{yint}) to be applied. In such a situation, the explicit $x \neq \xi$ dependent Shuvaev transform (\ref{eq:shuvg}) may be used, with the $x,x'$ given by (\ref{xxp}), although formally this corresponds to a higher order contribution, being driven by the $k_\perp^2 \sim M_V^2$ region, as well as depending on the details of the meson wave function. Thus some care may be needed when applying the results presented in Table~\ref{pdftjpsi}. On the other hand, for $J/\psi$ production at higher photon $Q^2$, as well as $\Upsilon$ production, where we are safely in the perturbative, $x' \ll x$, regime, this is much less of an issue. In this case, the importance of such a careful treatment is clearer, in particular when making comparison with high precision data.

\section{Conclusions}\label{sec:conc}

Generalized parton distributions (GPDFs) are a crucial ingredient in a wide range of hadronic processes, from deeply virtual Compton scattering ($\gamma^* p \to \gamma p$),  to diffractive vector particle production ($\gamma^{(*)} p \to V p$ where $V=\rho, J/\psi, \Upsilon, Z$...) and central exclusive production \linebreak[4]($pp \to p\, +\,X\,+\,p$, where $X=$ Higgs particle, dijets, $\chi_c$...). However, as the cross sections for such exclusive processes are generally small, there are insufficient data to determine these multi--argument distributions with an accuracy comparable to that of the global parton analyses for the diagonal distributions. The Shuvaev transform~\cite{Shuvaev:1999fm,Shuvaev:1999ce} offers a way to avoid this problem, by relating the GPDFs to the diagonal distributions in the $\xi \ll 1$ region. 

In this paper we have demonstrated that the Shuvaev transform for the quark and gluon GPDFs can be recast in a particularly simple form when $x = \xi \ll 1$. This allows for an easy way to bypass the poorly convergent standard form of the transform, without making any assumptions about the low--$x$ behaviour of the diagonal PDFs. For illustration we have considered the specific cases of Higgs boson central exclusive production and $J/\psi$, $\Upsilon$ photoproduction at the LHC. We have shown how the simple form of (\ref{yint}) can be used to give a precise treatment of the scale dependence of the unintegrated gluon GPDFs, and have observed that the predicted cross sections, in particular for Higgs boson CEP, can be somewhat underestimated by a common approximation for the Shuvaev transform in the low--$x$ region, which our treatment avoids. We have also found a similar, although less pronounced, result in the case of $J/\psi$ and $\Upsilon$ photoproduction. 

Our simple results, given by (\ref{yint}) and (\ref{yintq}), allow for a more precise and readily implemented form of the Shuvaev transform in the $x \approx \xi \ll 1$ region, avoiding any further approximations, and should have a wide range of phenomenological applications for exclusive and diffractive processes.

\section*{Acknowledgements}

The author is grateful to Jeff Forshaw, Valery Khoze and Misha Ryskin for useful discussions and for bringing this issue to his attention.

\bibliography{ggbib}{}

\begin{thebibliography}{10}

\bibitem{Diehl:2003ny}
M.~Diehl,
\newblock Phys.Rept. {\bf 388}, 41 (2003), hep-ph/0307382.

\bibitem{Belitsky:2005qn}
A.~Belitsky and A.~Radyushkin,
\newblock Phys.Rept. {\bf 418}, 1 (2005), hep-ph/0504030.

\bibitem{Kumericki:2009uq}
K.~Kumericki and D.~Mueller,
\newblock Nucl.Phys. {\bf B841}, 1 (2010), 0904.0458.

\bibitem{Sabatie:2012pe}
F.~Sabatie and H.~Moutarde,
\newblock PoS {\bf QNP2012}, 016 (2012), 1207.4655.

\bibitem{Kroll:2012sm}
P.~Kroll, H.~Moutarde, and F.~Sabatie,
\newblock Eur.Phys.J. {\bf C73}, 2278 (2013), 1210.6975.

\bibitem{Shuvaev:1999fm}
A.~Shuvaev,
\newblock Phys.Rev. {\bf D60}, 116005 (1999), hep-ph/9902318.

\bibitem{Shuvaev:1999ce}
A.~Shuvaev, K.~J. Golec-Biernat, A.~D. Martin, and M.~Ryskin,
\newblock Phys.Rev. {\bf D60}, 014015 (1999), hep-ph/9902410.

\bibitem{Mueller:1998fv}
D.~Mueller, D.~Robaschik, B.~Geyer, F.~Dittes, and J.~Horejsi,
\newblock Fortsch.Phys. {\bf 42}, 101 (1994), hep-ph/9812448.

\bibitem{Musatov:1999xp}
I.~Musatov and A.~Radyushkin,
\newblock Phys.Rev. {\bf D61}, 074027 (2000), hep-ph/9905376.

\bibitem{Radyushkin:2000uy}
A.~Radyushkin,
\newblock (2000), hep-ph/0101225.

\bibitem{Kimber:1999xc}
M.~Kimber, A.~D. Martin, and M.~Ryskin,
\newblock Eur.Phys.J. {\bf C12}, 655 (2000), hep-ph/9911379.

\bibitem{Martin:2001ms}
A.~D. Martin and M.~Ryskin,
\newblock Phys.Rev. {\bf D64}, 094017 (2001), hep-ph/0107149.

\bibitem{Khoze:2000cy}
V.~A. Khoze, A.~D. Martin, and M.~Ryskin,
\newblock Eur.Phys.J. {\bf C14}, 525 (2000), hep-ph/0002072.

\bibitem{Ivanov:2004ax}
I.~Ivanov, N.~Nikolaev, and A.~Savin,
\newblock Phys.Part.Nucl. {\bf 37}, 1 (2006), hep-ph/0501034.

\bibitem{Kowalski:2006hc}
H.~Kowalski, L.~Motyka, and G.~Watt,
\newblock Phys.Rev. {\bf D74}, 074016 (2006), hep-ph/0606272.

\bibitem{Martin:2007sb}
A.~Martin, C.~Nockles, M.~G. Ryskin, and T.~Teubner,
\newblock Phys.Lett. {\bf B662}, 252 (2008), 0709.4406.

\bibitem{Maciula:2011iv}
R.~Maciula, R.~Pasechnik, and A.~Szczurek,
\newblock Phys.Rev. {\bf D84}, 114014 (2011), 1109.5517.

\bibitem{Jones:2013pga}
S.~Jones, A.~Martin, M.~Ryskin, and T.~Teubner,
\newblock (2013), 1307.7099.

\bibitem{Ji:1996ek}
X.-D. Ji,
\newblock Phys.Rev.Lett. {\bf 78}, 610 (1997), hep-ph/9603249.

\bibitem{Ji:1996nm}
X.-D. Ji,
\newblock Phys.Rev. {\bf D55}, 7114 (1997), hep-ph/9609381.

\bibitem{Ji:1998pc}
X.-D. Ji,
\newblock J.Phys. {\bf G24}, 1181 (1998), hep-ph/9807358.

\bibitem{Radyushkin:1997ki}
A.~Radyushkin,
\newblock Phys.Rev. {\bf D56}, 5524 (1997), hep-ph/9704207.

\bibitem{GolecBiernat:1998ja}
K.~J. Golec-Biernat and A.~D. Martin,
\newblock Phys.Rev. {\bf D59}, 014029 (1999), hep-ph/9807497.

\bibitem{Ohrndorf:1981qv}
T.~Ohrndorf,
\newblock Nucl.Phys. {\bf B198}, 26 (1982).

\bibitem{Diehl:2007zu}
M.~Diehl and W.~Kugler,
\newblock Phys.Lett. {\bf B660}, 202 (2008), 0711.2184.

\bibitem{Martin:2009zzb}
A.~D. Martin, C.~Nockles, M.~G. Ryskin, A.~G. Shuvaev, and T.~Teubner,
\newblock Eur.Phys.J. {\bf C63}, 57 (2009).

\bibitem{rginter}
{\tt http://www.maths.liv.ac.uk/TheorPhys/RESEARCH/pubcodes.html}.

\bibitem{Martin:2009iq}
A.~D. Martin, W.~J. Stirling, R.~S. Thorne, and G.~Watt,
\newblock Eur.Phys.J. {\bf C63}, 189 (2009), 0901.0002.

\bibitem{Khoze:2001xm}
V.~A. Khoze, A.~D. Martin, and M.~G. Ryskin,
\newblock Eur.Phys.J. {\bf C23}, 311 (2002), hep-ph/0111078.

\bibitem{Albrow:2010yb}
M.~G. Albrow, T.~D. Coughlin, and J.~R. Forshaw,
\newblock Prog.Part.Nucl.Phys. {\bf 65}, 149 (2010), 1006.1289.

\bibitem{HarlandLang:2013jf}
L.~A. Harland-Lang, V.~A. Khoze, M.~G. Ryskin, and W.~J. Stirling,
\newblock (2013), 1301.2552.

\bibitem{Khoze:1997dr}
V.~A. Khoze, A.~D. Martin, and M.~Ryskin,
\newblock Phys.Lett. {\bf B401}, 330 (1997), hep-ph/9701419.

\bibitem{Khoze:2004yb}
V.~A. Khoze, A.~D. Martin, M.~G. Ryskin, and W.~J. Stirling,
\newblock Eur.Phys.J. {\bf C35}, 211 (2004), hep-ph/0403218.

\bibitem{HarlandLang:2012qz}
L.~A. Harland-Lang, V.~A. Khoze, M.~G. Ryskin, and W.~J. Stirling,
\newblock Eur.Phys.J. {\bf C72}, 2110 (2012), 1204.4803.

\bibitem{Coughlin:2009tr}
T.~Coughlin and J.~Forshaw,
\newblock JHEP {\bf 1001}, 121 (2010), 0912.3280.

\bibitem{Martin:1997wy}
A.~D. Martin and M.~Ryskin,
\newblock Phys.Rev. {\bf D57}, 6692 (1998), hep-ph/9711371.

\bibitem{HarlandLang:2010ep}
L.~A. Harland-Lang, V.~A. Khoze, M.~G. Ryskin, and W.~J. Stirling,
\newblock Eur.Phys.J. {\bf C69}, 179 (2010), 1005.0695.

\bibitem{Pumplin:2002vw}
J.~Pumplin {\em et~al.},
\newblock JHEP {\bf 0207}, 012 (2002), hep-ph/0201195.

\bibitem{Lai:2010vv}
H.-L. Lai {\em et~al.},
\newblock Phys.Rev. {\bf D82}, 074024 (2010), 1007.2241.

\bibitem{Gluck:2007ck}
M.~Gluck, P.~Jimenez-Delgado, and E.~Reya,
\newblock Eur.Phys.J. {\bf C53}, 355 (2008), 0709.0614.

\bibitem{Ball:2010de}
R.~D. Ball {\em et~al.},
\newblock Nucl.Phys. {\bf B838}, 136 (2010), 1002.4407.

\bibitem{Ryskin:1992ui}
M.~Ryskin,
\newblock Z.Phys. {\bf C57}, 89 (1993).

\bibitem{Ryskin:1995hz}
M.~Ryskin, R.~Roberts, A.~D. Martin, and E.~Levin,
\newblock Z.Phys. {\bf C76}, 231 (1997), hep-ph/9511228.

\bibitem{Martin:1999rn}
A.~D. Martin, M.~Ryskin, and T.~Teubner,
\newblock Phys.Lett. {\bf B454}, 339 (1999), hep-ph/9901420.

\bibitem{Martin:1999wb}
A.~D. Martin, M.~Ryskin, and T.~Teubner,
\newblock Phys.Rev. {\bf D62}, 014022 (2000), hep-ph/9912551.

\end{thebibliography}
\bibliographystyle{h-physrev}

\end{document}